\documentclass[%
 aip,
 amsmath,amssymb,
 reprint,%
]{revtex4-1}

\usepackage{graphicx}
\usepackage{dcolumn}
\usepackage{bm}
\usepackage{color}

\usepackage[utf8]{inputenc}
\usepackage[T1]{fontenc}
\usepackage{mathptmx}

\graphicspath{ {./figures/} }

\begin{document}

\preprint{AIP/123-QED}

\title[Quantum turbulence in Bose-Einstein condensates: present status and new challenges ahead]{Quantum turbulence in Bose-Einstein condensates: present status and new challenges ahead}

\author{L. Madeira}
\email{madeira@ifsc.usp.br}
\affiliation{Instituto de F\'isica de S\~ao Carlos, Universidade de S\~ao Paulo, CP 369, S\~ao Carlos, S\~ao Paulo, Brazil}

\author{A. Cidrim}
\affiliation{Departamento de F\'isica, Universidade Federal de S\~ao Carlos, S\~ao Carlos, S\~ao Paulo, 13565-905, Brazil}

\author{M. Hemmerling}
\affiliation{Instituto de F\'isica de S\~ao Carlos, Universidade de S\~ao Paulo, CP 369, S\~ao Carlos, S\~ao Paulo, Brazil}

\author{M.A. Caracanhas}
\affiliation{Instituto de F\'isica de S\~ao Carlos, Universidade de S\~ao Paulo, CP 369, S\~ao Carlos, S\~ao Paulo, Brazil}

\author{F.E.A. dos Santos}
\affiliation{Departamento de F\'isica, Universidade Federal de S\~ao Carlos, S\~ao Carlos, S\~ao Paulo, 13565-905, Brazil}

\author{V.S. Bagnato}%
\affiliation{Instituto de F\'isica de S\~ao Carlos, Universidade de S\~ao Paulo, CP 369, S\~ao Carlos, S\~ao Paulo, Brazil}
\affiliation{
Hagler Fellow, Department of Biomedical Engineering, Texas A\&M University, College Station, Texas 77843, USA
}%

\date{\today}

\begin{abstract}
The field of quantum turbulence is related to the manifestation of turbulence in quantum fluids, such as liquid helium and ultracold gases.
The concept of turbulence in quantum systems was conceived
more than 70 years ago by Onsager and Feynman, but the study of turbulent ultracold gases is very recent.
Although it is a young field, it already provides new approaches to the problem of turbulence.
We review the advances and present status, of both theory and experiments, concerning atomic Bose-Einstein condensates (BECs).
We present the difficulties of characterizing turbulence in trapped BECs, if compared to classical turbulence or turbulence in liquid helium.
We summarize the challenges ahead, mostly related to the understanding of fundamental properties of quantum turbulence, including what is being done to investigate them.
\end{abstract}

\maketitle

\section{Introduction}

Turbulence is characterized by a large number of degrees of freedom
interacting nonlinearly, over several length scales, to
produce a disordered state both in space and time.
From a technical point of view, turbulence is considered a harmful or highly undesirable process, taking as an example flow of gas through pipes or a motion of a piston in an engine which, due to wall friction, produces great energy losses.
However, turbulence is present on almost all levels of the organization of matter, ranging from quantum to astrophysical scales.

Classical turbulence (CT) has been studied for much longer than its quantum version, quantum turbulence (QT).
Since there are many aspects of CT that are not well-understood, dealing with QT may seem a formidable challenge.
However, the vortex circulation
is quantized in QT and continuous in CT, so the quantum version may be more tractable. One of the main questions that needs to be answered is until what extend our knowledge on classical turbulence will help to explain turbulence in quantum systems \cite{barenghi82,nazarenko11book}.

Bose-Einstein condensates (BECs) are excellent candidates for the study of QT due to the amount of control that is possible in these systems\cite{inouye98,goerlitz01,bloch08,henderson09,white14}. Interparticle interactions can be tuned via Feshbach resonances, and different trapping geometries have been successfully employed.
However, there are also some intrinsic difficulties in these systems, mostly related to the small range of length scales available and visualization techniques.
Despite these intrinsic difficulties, much has
been accomplished \cite{tsubota08,tsatos16,madeira20}.
One of the most important milestones since the first observation of turbulence in a trapped BEC, and its signature self-similar expansion \cite{henn09}, was the observation of a
power-law in the energy spectrum \cite{seman11,thompson13,navon16}.

In this review, we emphasize QT in trapped BECs. We discuss CT and QT in liquid helium in order to be able to compare and contrast QT in trapped dilute atomic BECs to these other systems.
This work is structured as follows. In Sec.~\ref{sec:pre} we introduce briefly topics and concepts that will be relevant throughout the manuscript, such as superfluidity in helium, quantized vortices, fundamentals of Bose-Einstein condensation, and how to excite BECs.
Aspects of classical turbulence,
specially those that we compare to QT,
are presented in Sec.~\ref{sec:classical}.
In Sec.~\ref{sec:quantum} we discuss the different types of quantum turbulence.
Examples of quantum turbulence in systems that are not single-component BECs are given in Sec.~\ref{sec:experiments}.
Lastly, we summarize the challenges that the field currently faces in Sec.~\ref{sec:challenges}.

\section{Preliminaries}
\label{sec:pre}

\subsection{Superfluid Helium}

Before the experimental realization of trapped Bose-Einstein condensates, quantum turbulence was already known and studied using liquid Helium.
For temperatures above 4K, $^4$He is a conventional gas. For temperatures below this value, unlike any other substance, it becomes liquid and remains in this state until absolute zero. This exceptional property owns to the low mass of Helium, which makes its zero-point energy high enough to suppress any process of crystallization \cite{pethick02} (unless a high pressure is applied).

Liquid He possesses two distinct phases separated by
a critical temperature point, $T_{\lambda}=$2.17K, known as lambda point due to the shape of the specific heat curve. To distinguish both phases,
$^4$He is called He I above $T_{\lambda}$, and He II below it.
Under certain conditions, Helium II exhibits unique properties such as zero entropy, zero viscosity, and superfluidity. The last refers to its ability to flow without friction.
This picture is helpful in many aspects, but idealized. In real experiments thermal noise will always be present.

Early experiments on the superfluidity of He II gave contradicting results.
One of the techniques demonstrating its frictionless nature was to determine Helium II viscosity from its flow through a long, narrow capillary.
The experimental results showed that below a critical velocity ($v_c$), Helium II indeed flows without friction \cite{kaptiza38}.
Similar results have been obtained using a rotating, torus-shaped apparatus filled with a porous material within which He II can flow freely \cite{reppy65}.
Initially, a fast-rotating torus is suddenly stopped allowing Helium to decrease its velocity slowly. Once the velocity reaches $v_c$, the superfluid begins flowing without dissipation.

On the other hand, these two examples were contradicted by the so-called Couette flow \cite{snyder74,tagg92,egbers00}.
The Couette flow experiment consisted of two cylinders: an outer rotating  one and an inner fixed one. The measurement of the torque exerted by the flow on the stationary cylinder is proportional to the viscosity of the fluid, which, unexpectedly, turned out to be non-zero. The reason for such conflicting results lies in the fact that Helium II has two components: normal fluid and superfluid, and their ratio is strongly temperature-dependent, meaning that only at $T$=0 He II is 100$\%$ superfluid \cite{barenghi01}.

Above the critical velocity, Helium II can sustain vorticity that, unlike classical vortices, can take only a filament form. Moreover, the angular momentum of the vortex is quantized, so its value is an integer number. The consequence of this is that rotating the superfluid faster than a critical value breaks it into several smaller discrete filaments creating an array of vortices. Quantum vortices have no classical counterpart in nature as they reach unusual dimensions for a single filament that can be as long as $10$ cm with a core radius of only $10^{-8}$ cm \cite{mccormack12}. Individual filaments or their arrays are the result of superfluid laminar flow. More common is its disordered spatial form called vortex tangle, which represents the turbulence of the superfluid component \cite{allen38,barenghi82,barenghi01}.

The concept of the two fluid model was first laid by Tisza \cite{tisza38} and mathematically formalized by Landau \cite{landau65,balibar07,balibar17}. Although theoretically complete, the two-fluid theory failed to explain the non-zero vorticity of the superfluid, which was demonstrated experimentally.
In fact, the model assumes \textit{a priori} zero-vorticity, defined as $\mathbf{\omega} = \nabla \times \mathbf{v}$, by arguing that the dissipation-free zero-viscosity fluid velocity field should be conservative. Hence, it could be written as $\nabla \phi = \mathbf{v}$.
Since $\mathbf{\omega} = \nabla \times \nabla \phi = 0$, then $\nabla \times \mathbf{v} = 0$, meaning the superfluid should be irrotational.

\subsection{Quantized vortices}\label{SubSec:Quantized_Vortices}

Landau's condition that the superfluid fraction of the He II is irrotational, $\nabla \times \mathbf{v} = 0$, can be closer examined following the argument of London who first connected BEC with superfluid He. London realized that the condensed atoms can be described by a macroscopic wave function if the collective occupation corresponds to the lowest energy state. For $N$ condensed atoms one could write $\Psi(\mathbf{r}, t) =\sqrt{\rho(\mathbf{r},t)} \, \exp\,[iS(\mathbf{r}, t )]$, where $\rho (\mathbf{r},t) = |\Psi (\mathbf{r}, t)| ^2$ is the density of the condensate and $S(\mathbf{r},t)$ its phase. Hence the probability current is $\mathbf{j} = \hbar / (2 mi) (\Psi^{*} \nabla \Psi - \Psi \nabla \Psi^{*}) = \rho(\hbar/m)\nabla S$. The above can be seen as a flux of the density that flows with velocity $\mathbf{v} = (\hbar/m)\nabla S$, which can be simplified to $\mathbf{j}=\rho \, \mathbf{v}$. The direct consequence of this is that when $S$ has continuous first and second derivatives, the velocity field is irrotational, $\nabla \times \mathbf{v} = 0$. However, this will not hold in the presence of the vortex line.

In 1949 Onsager
suggested that the circulation can be quantized \cite{eyink06}. For given path $C$ the circulation is $\Gamma = \oint_C  \mathbf{v} \cdot d \mathbf{r}$ . Hence, for the single-valued macroscopic wave function, with the $2\pi n$ phase change ($n$ is an integer), the circulation can be written as 
\begin{equation}
\Gamma = \oint_C \mathbf{v} \cdot d \mathbf{r} = \frac{\hbar}{m} 2 \pi n = n \kappa .
\end{equation}
The integer $n$ is often called the charge of the vortex.
In most cases, quantized vortex lines can be thought as an ordinary (classical) vortex line in the superfluid with the hollow core and the quantized circulation $\kappa$, ($\hbar$ is Planck's constant and $m$ is the mass of He).

Following Osanger's proposal, in 1950 Osborn found that the rotation of the Helium II was indistinguishable from the rotation of Helium I \cite{Osborne_1950}. Later, in 1955, Feynman presented his theory on vortex lines \cite{feynman55}, which finally led to the conclusion that when He II is spun in the cylinder, the normal fluid rotates uniformly with it, while the superfluid forms discrete vortex filaments aligned parallel to the cylinder axis\cite{donnelly_swanson_1986}. 

The angular momentum of a vortex is quantized and proportional to its charge. In principle, a superfluid with high angular momentum could be in a state of a single vortex with a large charge $Q$, or $Q$ single-charged vortices. However, the vortex kinetic energy is proportional to $Q^2$, and a state with one vortex with a large charge will not be energetically favourable \cite{barenghi16}. This has been confirmed experimentally \cite{shin04}.

\subsection{Bose-Einstein Condensation}

BECs are great toolboxes for studying numerous fields of Physics, ranging from statistical mechanics to field theories.
Their strongest advantage is that they allow the investigation of quantum effects on a macroscopic scale. 
BEC relies on the indistinguishability and wave nature of particles. The condensation itself is a phase transition process occurring when a wavepacket of a boson reaches the size of the thermal de Broglie wavelength ($\lambda_{dB} \propto n^{-1/3}$, where $n$ is the interparticle density) \cite{pethick02}.
The whole gas of bosons occupies the lowest energy level by constructive interference between the individual wavepackets. The conditions for the condensation are a critical density and a critical temperature.
Most condensates are obtained in a nonhomogenous harmonic trap for which the critical temperature depends on only two properties: the number of atoms $N$ and the trap frequencies. The critical temperature is given by $T_c = 0.15 \bar{\omega} N^{1/3}/k_B $, where $\bar{\omega}=(\omega_x\omega_y\omega_z)^{1/3}$ is the geometrical mean of the three trapping frequencies. 
To reach the desired temperature and density, a combination of magnetic fields and optical forces is used. The final stage requires removing the hottest atoms from the system, which is called the evaporation technique \cite{ketterle99}.

The existence of turbulence in trapped BECs allows us to study the properties and the dynamics of the condensate itself, and to explore some of the universal characteristics related to turbulence.

\subsection{Exciting a BEC}

Vortices and turbulence in BECs have been generated by moving a laser beam across the BEC \cite{raman2001,neely2010,white2012,stagg2015}, shaking it \cite{henn09}, rotating the trap around two perpendicular axes \cite{tsubota02}, phase imprinting staggered vortices \cite{white10}, or by thermally quenching the system (Kibble-Zurek mechanism) \cite{weiler2008,lamporesi13,navon2015}. The turbulent states resulting from this broad range of techniques, however, are hard to compare. Moreover, some of them bring additional excitations and fragmentation of the condensate \cite{mueller06}, which makes it impossible to obtain a clean turbulent regime.

The decay of multicharged vortices gives an alternative scenario to induce turbulence in BECs. Multiply quantized vortices are energetically unstable and decay into singly quantized vortices \cite{shin04,kumakura2006,isoshima2007}. Besides the energy instability, there is also a dynamical instability \cite{pu1999,mottonen2003,huhtamaki2006}, which can destabilize the vortices even in the absence of dissipation, at zero temperature. A controlled technique for the generation of multicharged vortices in atomic condensates is the technique of topological phase imprinting \cite{nakahara2000,leanhardt2002,shin04}.

\section{Classical turbulence}
\label{sec:classical}

Turbulence is the manifestation of the spatial-temporal chaotic behavior of the fluid flows at large Reynolds numbers, i.e., of a strongly nonlinear dissipative system with a vast number of degrees of freedom (most probably) described by the Navier Stokes equation (NSE) \cite{foias01}.
Since it is an extremely complicated phenomenon, it is challenging to give a precise definition of turbulence. Below we list some of the characteristics of classical turbulent flows: 

$\bullet$  Large Reynolds numbers: an important characteristic of a viscous flow is its Reynolds number ($Re$), which is defined as $Re=\frac{L V }{\nu}$, with $L$ and $V$ being respectively a characteristic scale and velocity of the flow, and $\nu$ its viscosity. Turbulence often originates as instability of laminar flow if $Re$ becomes too large.  The instability is related to the interaction of viscous terms and nonlinear inertia terms in the equations of motion.

$\bullet$ Wide range of strongly interacting scales: the interaction between the many degrees of freedom results from the nonlinearity of turbulent flows.

$\bullet$ Dissipation: turbulent flows are highly dissipative. A source of energy is required to maintain turbulence. Typically, in three-dimensions, the energy supply is at large length scales and its dissipation at small ones.

$\bullet$ Diffusivity: it is associated with the strongly enhanced transport processes in turbulence. It causes rapid mixing and increased rates of momentum, heat, and mass transfer.

$\bullet$ Intrinsic spatio-temporal randomness: there is no necessity for external random forcing, provided Reynolds number is large enough.

$\bullet$ Vorticity fluctuations: high levels of fluctuating vorticity characterize turbulence. For this reason, vorticity plays an essential role in the description of turbulent flows.

$\bullet$ Turbulence is not a feature of fluids, but of fluid flows. Most of the dynamics of turbulent flows are the same in all fluids, regardless of their molecular properties, if their $Re$ is large enough. 

Randomness and nonlinearity combine to make the equation of turbulence theory suffer from the absence of sufficiently robust mathematical methods \cite{lumley72}. Since the equation of motion is nonlinear, each individual flow pattern has specific unique characteristics that are associated with its initial and boundary conditions. No general solution of the NSE is known. However, no turbulence theory intends to deal with all kinds and types of flows in a general way. Instead, theoreticians concentrate on families of flows with relatively simple boundary conditions.

\subsubsection{Navier Stokes}

Consider the incompressible (constant density) Navier-Stokes equation,
\begin{eqnarray}\label{NSE}
 &&\partial_t\mathbf{v}+\mathbf{v}\cdot\nabla\mathbf{v}=-\frac{1}{m \rho}\nabla p+\nu\nabla^2\mathbf{v},\\
  &&\nabla\cdot\mathbf{v}=0,
\end{eqnarray} where $\nu$ is the kinematic viscosity, $p$ is the pressure, $m$ and $\rho$ are particle mass and density, respectively. The NSE is the result of coarse graining over the stochastic molecular effects. Although the NSE has a limited kinetic foundation, it is commonly believed to be adequate to describe real fluid flows.

The mathematics of nonlinear partial differential equations has not been developed to a point where a general
solution can be given, thus there is little substantial analytical use of the NSE in turbulence.
However, numerical simulations have been used extensively.
Furthermore, the NSE is non-integrable and nonlocal.
The property of nonlocality in physical space is due to the pressure,
which is directly defined by the velocity field.
Hence the velocity field is nonlocal in physical and any other space (there is a coupling between large and small length scales). The difficulties described above are mostly of a formal or technical nature. There is another difficulty of a general nature: the lack of knowledge about the underlying physical processes of turbulence and its generation and origin \cite{homes12} .

Formally, the problem of the NSE subject to initial and boundary conditions can be solved numerically. However, looking at the behavior of a particular solution, which does not have analytical form, does not solve the problem since any particular solution may not contribute to the understanding of the fundamental physics of turbulent flows. We need an understanding of the global behavior of the system, and all NSE solutions to elucidate the phenomenon of turbulence.

Just like in statistical physics, the statistical approach should be adopted in turbulence theories from the outset start due to the extreme complexity. In turbulence, however, the equations of motion always lead to situations in which there are more unknowns than equations. This is called the closure problem of turbulence theory.
One has to use statistical assumptions to make the number of equations equal to the number
of unknowns, as will be illustrated in the following section.
 
Symmetry considerations are central to the study of fully developed turbulence \cite{foias01}. At higher $Re$, when the flow becomes turbulent, its statistical properties are invariant under translations. Similar remarks can be made about all the other symmetries of the NSE: the mechanism responsible for the generation of the turbulent flow are usually not consistent with most of the possible symmetries. However, the qualitative aspects of many turbulent flows suggests some form of homogeneity, isotropy, and possible scale-invariance.
In the limit of infinite $Re$, all possible symmetries of the NSE, usually broken by the mechanism producing the turbulent flow, are restored in a statistical sense at small scales ($\ell \ll L$, $L$ being the scale characteristic of the production of turbulence). Under the same assumption, the turbulent flow is self-similar at small scales, i.e., it possesses a unique scaling exponent $h$. Thus, there exists a scaling exponent $h\in \mathbb{R}$ such that $\delta\mathbf{v}(\mathbf{r},\lambda\mathbf{\ell}) = \lambda^h \delta\mathbf{v}(\mathbf{r},\mathbf{\ell})$ for all $\lambda\in\mathbb{R}_+$, for all $\mathbf{r}$ and all increments $\mathbf{\ell}$ and $\lambda\mathbf{\ell}$ small compared to $L$.

Based on these symmetry properties of the NSE described before, phenomenological theories of turbulence can be applied to make crucial assumptions in the early stage of the analysis. In many circumstances, it is possible to argue that some aspects of the structure of turbulence depend only on a few independent variables or parameters. If such a situation prevails, dimensional methods often dictate the relationship between the dependent and independent variables, which results in a solution that is known except for a numerical coefficient. Another frequently used approach is to exploit some of the asymptotic properties of turbulent flows. Any proposed descriptions of turbulence should behave appropriately in the limit where the $Re$ approaches infinity since one can consider vanishing small effects of the molecular viscosity. Also, in simple flow geometries, the characteristics of the turbulent motion at some point in time and space appear to be controlled mainly by the immediate environment (local invariance).

\subsubsection{Kolmogorov}

Though there exist a set of deterministic differential equations (NSE) probably containing almost all of turbulence, most of our knowledge about turbulence comes from observations and experiments.  Phenomenology is the most commonly used description for some statistical aspects of turbulent flows, since it is based on or motivated by some experimental data. It involves the use of dimensional analysis, a variety of scaling arguments, symmetries, invariant proprieties, and various assumptions, some of which are of unknown validity and obscure physical and mathematical justification \cite{tsinober07}. The famous Kolmogorov hypotheses belong to this category \cite{frisch95}. We start with what is called Kolmogorov phenomenology. We will see that the finite limit of the mean energy dissipation $\varepsilon$ at $Re\rightarrow\infty$ defines a unique scaling exponent in Kolmogorov's  $2/3$ law \cite{kolmogorov41a,kolmogorov41b}, while his $4/5$ law \cite{kolmogorov41c}, otherwise,  is a consequence of NSE.

In turbulent flows, a wide range of length scales exists bounded from above by the dimensions of the flow and bounded from below by the diffusive action of molecular viscosity. Incidentally, this is the reason why spectral analysis of turbulent motion is useful \cite{brachet00}. The parameters governing the small scale motion include at least the dissipation rate per unit mass $\varepsilon$ $\;[m^2 s^{-3}]$ and the kinematic viscosity $\nu$ $\;[m^2 s^{-1}]$. With these parameters, one can form length, time, and velocity scales as follows:
\begin{eqnarray} \nonumber
 \eta &\equiv& (\nu^3/\varepsilon)^{1/4},\\  \nonumber
 \tau&\equiv&(\nu/\varepsilon)^{1/2},\\
 \upsilon&\equiv&(\nu \varepsilon)^{1/4}.
 \end{eqnarray}
These scales are referred to as the Kolmogorov micro scales. The $Re$ formed with $\eta$ and $\upsilon$ is equal to one $\eta\;\upsilon/\nu =1$, which illustrates that the small-scale motion is quite viscous and that the viscous dissipation adjusts itself to the energy supply by adjusting length scales.  The small-scale motion should depend only on the rate at which it is supplied with energy by the large-scale motion and on the kinematic viscosity.

In the limit of infinite $Re$, all the small-scale statistical properties are uniquely and universally determined by the scale $\ell$ and the mean energy dissipation rate $\varepsilon$.  That is the Kolmogorov hypotheses of local isotropy, which postulates that a large $Re$ all the symmetries of the NSE are restored in the statistical sense. The possible scale invariance symmetries of NSE at $Re >> 1$ allowed to find $2/3$ famous exponent in the so-called inertial range of scales $r$, $L>>r>>\eta$. As an illustration of it, consider the second-order structure function $\langle(\delta\mathbf{v}(\ell))^2\rangle$. The straightforward dimensional analysis shows that this quantity has dimension $[m]^2[s]^{-2}$, where $[m]$ and $[s]$ are unit length and time, respectively. Since the mean energy dissipation rate per unit mass $\varepsilon$ has dimensions $[m]^2[s]^{-3}$, it follows from the universality assumption \cite{frisch95}

\begin{equation} \label{eqk1} \langle(\delta\mathbf{v}(\mathbf{r},\ell))^2\rangle = C \varepsilon^{2/3} \ell^{2/3},\end{equation} with $C$ is a universal dimensionless constant. It is noteworthy that Kolmogorov never worked in Fourier space. This was done by his PhD student who formulated the
 $-5/3$ law for the energy spectrum $E(k) = C_k  \varepsilon^{2/3} k^{-5/3}$, which is equivalent in some sense to Eq. \ref{eqk1}.

In the third 1941 paper \cite{kolmogorov41c}, Kolmogorov found that an exact relation can be derived for the third-order longitudinal structure-function. He assumed homogeneity, isotropy, and finiteness of energy dissipation. Without any further assumptions, he derived his $4/5$ law from the NSE. In the limit of infinite $Re$, the $\langle(\delta\mathbf{v}(\mathbf{r},\ell))^3\rangle$ of homogeneous isotropic turbulence evaluated for increments $\ell$ small compared to the integral scale is given in terms of the mean energy dissipation per unit mass $\varepsilon$ by  \cite{kolmogorov41a,kolmogorov41b,kolmogorov41c}
\begin{equation} \label{eqk2} \langle(\delta\mathbf{v}(\mathbf{r},\ell))^3\rangle= -\frac{4}{5}\varepsilon\ell. \end{equation} This is one of the most important results in fully developed turbulence because it is both exact and nontrivial. His $-4/5$ law obtained as a direct consequence of NSE for the inertial range $L>>r>>\eta$ . This relation, however,  was obtained for globally and not for locally isotropic turbulence. It is noteworthy that there is a considerable deviation from the $4/5$ law. In contrast, these same causes (lack of asymptotic, homogeneity, isotropy, finiteness of energy dissipation, and poor quality of data) have little effects on $2/3$ law. This seems surprising since the law in Eq.\ref{eqk2} is a consequence of the NSE, while Eq.\ref{eqk1} law is only a consequence of dimensional hypotheses. The $4/5$ law applies strictly to globally isotropic turbulent flows. Even with very large $Re$, the system may lack local isotropy.

The Kolmogorov papers raised several fundamental issues that have kept the turbulence community an active field until now. 

\subsubsection{Energy Cascade}

The cascade picture of turbulent flows takes its origin from Richardson \cite{richardson22,kolmogorov62}. The cascade picture is based on the intuitive notion that turbulent flows posses a hierarchical structure consisting of eddies as a result of successive instabilities.  The eddies of various sizes are represented as blobs stacked in decreasing sizes. The uppermost eddies have scales $L_0$. The successive generations of eddies have scales $L_n=L_0\; r^n (n=0,1,2,...)$, where $0<r<1$. The smallest eddies have scale $\sim \eta$, the Kolmogorov dissipative scale. The number of eddies per unit of volume is assumed to grow with $n$ as $r^{-3n}$ to ensure that small eddies are space-filling as large ones. Energy, introduced at the top at a rate $\varepsilon$, is cascading down this hierarchy of eddies at the same rate $\varepsilon$ and is eventually removed by dissipation at the bottom, still at the rate $\varepsilon$. The main advantage of the cascade picture is that it brings the scale-invariance phenomenology assumption of the Kolmogorov laws within the inertial range.

The Richardson-Kolmogorov cascade \cite{kolmogorov62} was formulated in physical space and is used frequently without much distinction both in physical and Fourier space. However, it was Neumann \cite{neumann49} who recognized that this process occurs not in physical space, but in Fourier space. That is, the nonlinear term in the NSE redistributes energy among the Fourier modes and not in scales as is frequently claimed (unless the scale is defined just as an inverse of the magnitude of the wavenumber of a Fourier mode). We recall that there is no contribution from the nonlinear term in the total energy balance equation since the nonlinear term in the energy equation has the form of a spatial flux. In other words, the nonlinear term redistributes the energy in physical space if the flow is statically non-homogeneous. Given the above arguments, it seems that the energy is dissipated not necessarily via a multi-step cascade-like process in physical space. Instead, there is an exchange of energy in both directions, whereas the dissipation occurs in small length scales.

\section{Quantum turbulence}
\label{sec:quantum}

Having as reference the characteristics of classical turbulence discussed before, one can start drawing comparisons with the dynamics of quantum fluids. The reason why turbulent phenonena should also appear in such systems is not only associated with the possibility of nucleating vortices in quantum fluids (see Sec.~\ref{SubSec:Quantized_Vortices}). For the particular case of atomic BECs, it is intimately related to the dynamical equation that successfully governs their dynamics, the Gross-Pitaevskii equation (GPE)\cite{gross61,pitaevskii61}, 
\begin{equation}
i \hbar \frac{\partial \Psi(\mathbf r,t)}{\partial t} = \left(-\frac{\hbar^2}{2m}\nabla^2 + V(\mathbf r) + g|\Psi(\mathbf r,t)|^2 \right)\Psi(\mathbf r,t), 
\label{Eq:GPE}
\end{equation}
which is simply a nonlinear Schr\"{o}dinger equation for the macroscropic wave-function $\psi$, with a generic potential $V$ and two-body contact interaction strength $g=4\pi a_s\hbar^2/m$ proportional to the $s$-wave scattering length $a_s$. Even though Eq.~\ref{Eq:GPE} is typically a rough approximation for describing the dynamics of superfluid helium (it is not valid for strongly interacting systems), for BECs it can accurately describe a myriad of regimes, in particular, turbulent phenomena. 

As discussed in Sec.~\ref{SubSec:Quantized_Vortices}, one can associate a velocity field $\mathbf{v}=(\hbar/m)\nabla S$ of a fluid, after considering the transformation $\Psi(\mathbf{r}, t) =\sqrt{\rho(\mathbf{r},t)} \, \exp\,[iS(\mathbf{r}, t )]$. The GPE can thus be rewritten in its hydrodynamical form
\begin{equation}
\frac{\partial\mathbf{v}}{\partial t} = -\frac{1}{m \rho} \nabla{}p - \nabla\left(\frac{v^2}{2}\right) + \frac{1}{m} \nabla\left(\frac{\hbar^2}{2m\sqrt{\rho}}\nabla^2\sqrt{\rho}\right) - \frac{1}{m}\nabla V,
\label{Eq:QuantumEuler}
\end{equation}
where a pressure-like term $p\equiv\rho^2 g/2$ is conveniently defined \cite{pethick08}. The only term proportional to $\hbar$ is the so-called \emph{quantum pressure}. It is a kinetic term that is only relevant when the density of the fluid varies abruptly in space. 

Notice that Eq.~\ref{Eq:QuantumEuler} closely resembles the NSE (Eq.~\ref{NSE}). In fact, in the limit $\hbar\rightarrow 0$, the above equation becomes identical to the \textit{Euler equation} for a classical irrotational fluid ($\nabla \times \mathbf v = 0$), which is the dissipation-free ($\nu=0$) form of the NSE. This indicates that the same nonlinearities present in the Euler equation, which one could argue that is the backbone of classical turbulence, are also present in quantum fluids that follow the GPE dynamics. By considering the multivalued nature of the field $S$ it is possible to correct Eq. (\ref{Eq:QuantumEuler}) in such a way that vorticity dynamics are also included\cite{dosSantos16}.

The mean-field description of the GPE is a $T=0$ model and ignores finite-temperature effects. In reality however atomic BECs are not composed entirely of a condensed fraction, which can be depleted by thermal collisions. Although these thermal effects can be sometimes negligible, one can attempt a more realistic description, taking thermal dissipation into account. A possible route for this more rigorous description is offered by the stochastic Gross-Pitaevskii theory~\cite{blakie2008,proukakis08}, which builds on the mean-field model, introducing coupling of low-lying modes to a reservoir of thermal modes. This is achieved by incorporating a dynamical noise to the GPE, which models quantum fluctuations, and also including a phenomenological dissipative effect from the thermal bath. This damping term $\gamma$ is added to the GPE\cite{tsubota02,proukakis08} by making $i \partial/\partial t \rightarrow ( i- \gamma)\partial/\partial t $. With this modification, the hydrodynamical form of the GPE (for an incompressible quantum fluid) corresponds then to a \emph{quantum Navier-Stokes} equation
\begin{equation}
\begin{aligned}
\frac{\partial\mathbf{v}}{\partial t} = & -\frac{1}{m\rho} \nabla{}p - \nabla\left(\frac{v^2}{2}\right) + \frac{1}{m} \nabla\left(\frac{\hbar^2}{2m\sqrt{\rho}}\nabla^2\sqrt{\rho}\right)\\ & - \frac{1}{m}\nabla V -\nu_q \nabla^2\mathbf v, 
\end{aligned}
\label{Eq:QuantumNS}
\end{equation}
where $\nu_q\equiv \hbar\gamma/2m$ is the analogous (quantum) kinematic viscosity \cite{bradley12}.  

When referring to turbulence in quantum fluids, we must be aware of the fact that it can be associated with a few types of disordered states. In the case of condensates, a turbulent state may be related to complex vortex dynamics or nonlinear interactions of waves. The latter is based on the general formalism of wave turbulence~\cite{nazarenko11book}, which has provided a framework to understand turbulence of wave-like excitations on structures such as: vortex lines (Kelvin waves), spin texture (see Sec.~\ref{sec:spin}), and also density fluctuations of a condensate. In this section we will focus on vortex turbulence and how it relates to its classical counterpart. 

It is crucial to highlight that even in a strict $T=0$ description, i.e. in the absence of thermal atoms, where dynamics is simply determined by the standard GPE, vortex turbulence finds an effective dissipative mechanism through \textit{vortex reconnections} \cite{wells15}. This means that, although the system is superfluid (and inviscid by its quantum nature), if one looks only to the rotation kinetic energy due to vortices, this can be lost in time and transformed into other forms of kinetic energy (related to density waves). This effective dissipation mechanism at $T=0$ shares close connection to the notion of an effective viscosity, first envisioned by Onsager \cite{onsager53} while noting the suggestive fact that a quantum of circulation $\hbar/m$ shares the same dimension as a kinematic viscosity. Following this idea, a superfluid Reynolds number was recently identified in a thorough numerical investigation of vortex-shedding dynamics in two dimensions\cite{reeves15}. 

\subsection{Vortex turbulence}

\subsubsection{Kolmogorov (or quasiclassical) turbulence}

Large-scale numerical simulations of the GPE describing the dynamics of randomly imprinted vortices in homogeneous BECs \cite{sasa11} have exemplified the emergence of a Kolmogorov-like energy spectrum of $k^{-5/3}$ in quantum systems. As in classical turbulence, this indicates the existence of an (incompressible) energy cascade from large to small length scales. The self-similarity encompassed by such scaling is believed to come from \textit{vortex bundling}. In analogy to the classical Richardson cascade, large bundles of vortices transfer energy to smaller bundles over many scales, down to typical lengths of a single quantum vortex (e.g. its core size). Ultimately, vortex-lines wiggle in Kelvin-wave motion, dissipating energy through sound emission \cite{skrbek12,barenghi14}. This cascade picture was further corroborated by numerical investigations using Biot-Savart models \cite{baggaley12}, which have also identified the temporal decay of the total vortex length $L\sim t^{-3/2}$ in this regime of bundling. However, although Kolmogorov energy spectra have been observed experimentally in superfluid helium \cite{maurer98,abid98}, there is yet no direct experimental observations of these vortex bundles. Due to its classical-like energy spectrum this regime is known as \textit{Kolmogorov} or \textit{quasiclassical quantum turbulence}. 

\subsubsection{Vinen (or ultraquantum) turbulence}

When the tangling vortices do not orient themselves in self-similar bundles throughout scales, the randomness in their orientations marks the absence of large-scale, energy-containing flow structure. As a consequence, the energy spectrum does not build up for small $k$ region and follows a $k^{-1}$ power-law \cite{baggaley12}, a scaling associated with a single quantum vortex energy spectrum. Due to its unique quantum nature, this different kind of turbulence is known as \textit{ultraquantum} or \textit{Vinen turbulence}, which has been investigated both experimentally \cite{vinen57,walmsley08} and numerically \cite{baggaley12b,stagg16,cidrim17}. Differently from Kolmogorov-type of turbulence, the Vinen regime follows a $L\sim t^{-1}$ temporal decay for the total vortex length \cite{vinen57}.

\begin{figure}[!htb]
\centering
\includegraphics[width=\columnwidth]{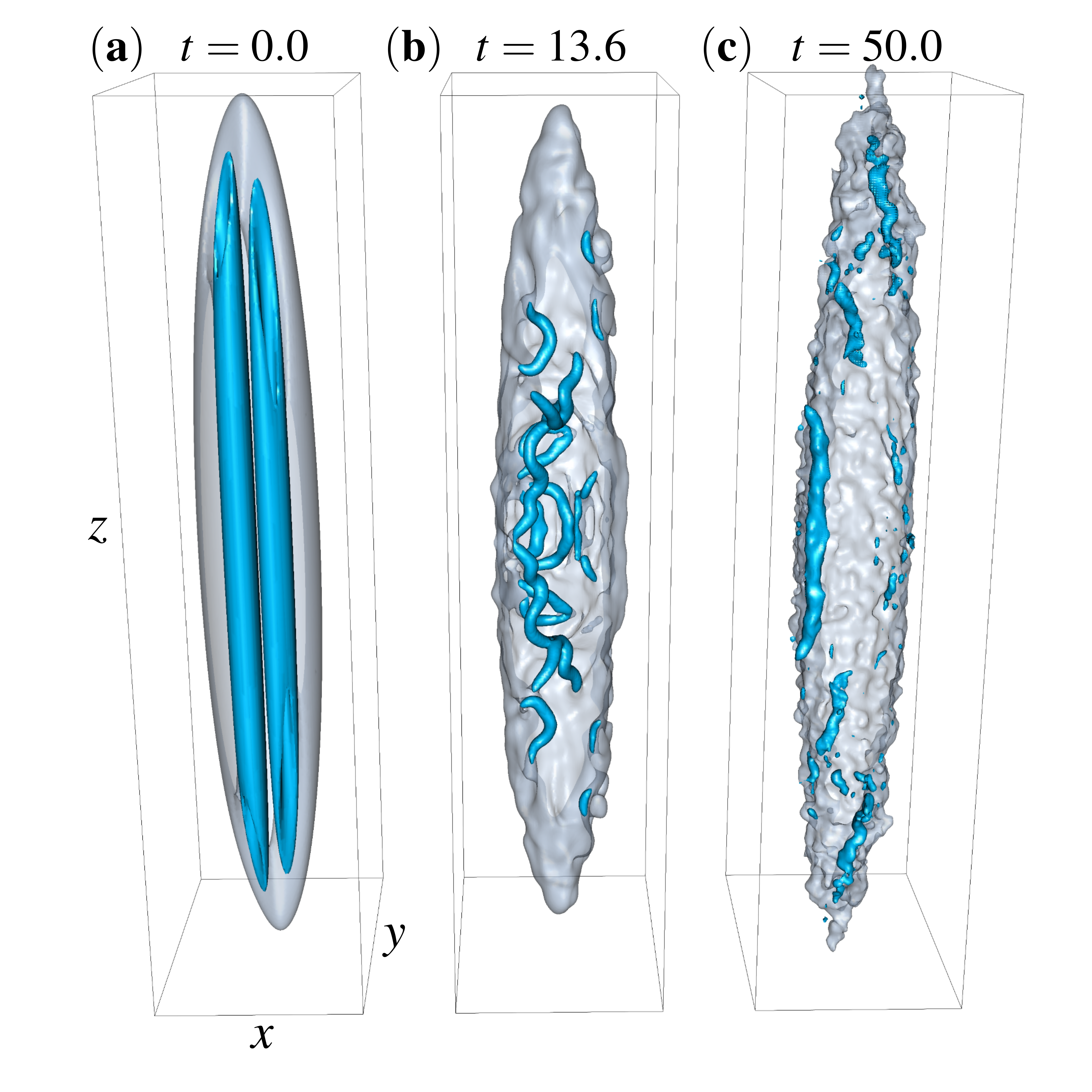} 
\caption{
Time evolution of the decay of two antiparallel doubly-charged vortices obtained via numerical simulations of the Gross-Pitaevskii equation.
It is possible to see the initial state with two imprinted vortices (a), which evolve to a turbulent quasi-isotropic state (b), and then decay (c).
Reprinted figure with permission from A. Cidrim, A. White, A. Allen,
V. S. Bagnato, and C. Barenghi, “Vinen turbulence via the decay of
multicharged vortices in trapped atomic Bose-Einstein
condensates,” Phys. Rev. A 96, 023617 (2017). Copyright 2017 by
the American Physical Society.
}
\label{fig:attis}
\end{figure}

In Reference \cite{cidrim17}, a turbulent state produced from the decay of multicharged vortices was investigated. The authors showed how the decay of multicharged vortices produced a turbulent state consistent with the Vinen regime of turbulence, see Fig.~\ref{fig:attis}. This was further corroborated by a particle and energy flux analysis \cite{Marino2020}. Additionally, non-Gaussian velocity statistics was also observed, a fundamental feature of quantum fluid turbulence and in distinction to the Gaussian statistics expected from classical turbulent systems\cite{white10,paoletti08}. 

\subsection{Scales availability and type of vortex turbulence}

Whether one can find a Kolmogorov or Vinen regime of vortex turbulence in a quantum fluid is firstly dependent on the ratio $L/\xi$, where $L$ is the typical largest scale of the system (e.g. size of a helium container or extension of the trap for an atomic superfluid) and $\xi$ is the healing length, which sets the approximate size of the vortex core. For a quantum system to exhibit the quasiclassical spectrum $k^{-5/3}$, a relatively large ratio is necessary, meaning few spatial decades, $\log (L/\xi)$, between $\xi$ and $L$ for the development of large-scale, self-similar structures. This requirement however is not necessary for the appearance of spectral features of ultraquantum turbulence. Experiments with superfluid He, for instance, typically offer $\log (L/\xi) \sim 4$ decades, justifying why both quasiclassical and ultraquantum limits have been observed \cite{vinen57,maurer98}. In contrast, in current experiments with atomic BECs~\cite{henn09,neely13,kwon14,seo17} $\log (L/\xi) \sim 1-2$, translating into system sizes which tend to hinder the formation of large-scale vortex bundles (or vortex clusters, in two dimensions). Therefore, these relatively small-ratio systems have only shown evidence of the Vinen turbulence regime. 

\subsection{3DQT}
\label{sec:experiments}

The high degree of control which is possible in weakly interacting dilute atomic BECs has motivated theoretical and experimental investigations of QT in these systems.
The S\~ao Carlos group reported the first evidences of QT in a trapped BEC \cite{henn09}. An external oscillatory perturbation combined with rotations was used to create a vortex tangle \cite{henn09,henn09_2}.
Different strengths and durations of the external oscillatory potential lead to different regimes \cite{seman11}.
Increasing the amplitude of oscillation first caused vortex nucleation and proliferation, then a
turbulent vortex regime,
and finally a
granulation of the condensate \cite{nguyen19}.

They also provided a way to identify a turbulent cloud. 
A thermal cloud is characterized by a gaussian density profile, while a BEC has a profile that reflects the shape of the confining trap, corresponding to the Thomas-Fermi regime.
For a cigar-shaped trap, the BEC cloud expands faster in the radial than in the axial direction, causing the inversion of the aspect-ratio during the free expansion.
In a turbulent BEC the situation is different.
For the cigar shape trap used in Ref.~\cite{henn09}, the turbulent BEC expands with a approximately constant aspect-ratio, following the release of the trap.
A generalized Lagrangian approach was devised to describe the dynamics of the cloud \cite{caracanhas13}.

The quantum behavior of BECs allows for interesting analogies.
Bose-Einstein condensates and atom lasers are both coherent matter-wave systems.
On the other hand, an optical speckle pattern is a random light map.
In Ref.~\cite{tavares17}, the authors
establish a parallel between a ground-state BEC with the propagation of an optical Gaussian beam, while the turbulent BEC was compared to an
elliptical speckle light map.
They showed that the correlations in BECs resemble the ones in
the Gaussian beams, while the same is true for the turbulent BECs and speckle
beams.
Hence, in principle, statistical atom optics could improve our understanding of quantum turbulence.

A discussion of relevant experimental realizations, but also of theoretical attempts to study the problem of quantum turbulence in quantum gases, has recently been compiled in Ref.~\cite{tsatos16}.
The firsts experiments involve a very limited number of settings, with the inhomogeneous density resulting from harmonic trapping bringing qualitative evidence for turbulence only, but no quantitative comparisons with theory.  

In references \cite{navon16,navon19} they eliminate this problem by studying turbulence in a weakly interacting homogeneous BEC, which was trapped in an optical box and driven out of equilibrium with an oscillating force. They observe the emergence of a turbulent cascade characterized by an isotropic power-law distribution in momentum space.
Its exponent was related to the weak-wave turbulence of a compressible superfluid. The same experimental apparatus was used to directly measure cascade fluxes in a turbulent system. As their system is thermally isolated from the environment, the dissipation occurs only in the form of particle loss. The optical box energy depth defines the particle and energy sink, which was controlled by changing the trapping laser power.
A tunable dissipation scale and a spatially uniform driven force allowed them to extract the cascade fluxes by studying the particles dissipation in the gas. In the understanding of quantum turbulence, the cascade fluxes are equally fundamental as the extensively studied steady-state power-law spectra, but much harder to be measured, which left them unexplored until now. These experiments establish the uniform Bose gas as a promising new platform for investigating many aspects of turbulence, as the interplay of vortex and wave turbulence by tuning the strength of nonlinearity via Feshbach resonances, and also the possibility to study the anisotropic turbulence by engineering arbitrary momentum-cutoff landscapes.

Experimental indicatives of vortex turbulence at small (healing length) scales were given by a piston shock experiment \cite{mossman18}. The generation of superfluid quantum turbulence arises as a consequence of the dissipation of excitations from the large scale shock front into small-scale vortex excitations, via a snake instability of a planar soliton train.

\begin{figure}[!htb]
\centering 
\includegraphics[width=0.8\columnwidth]{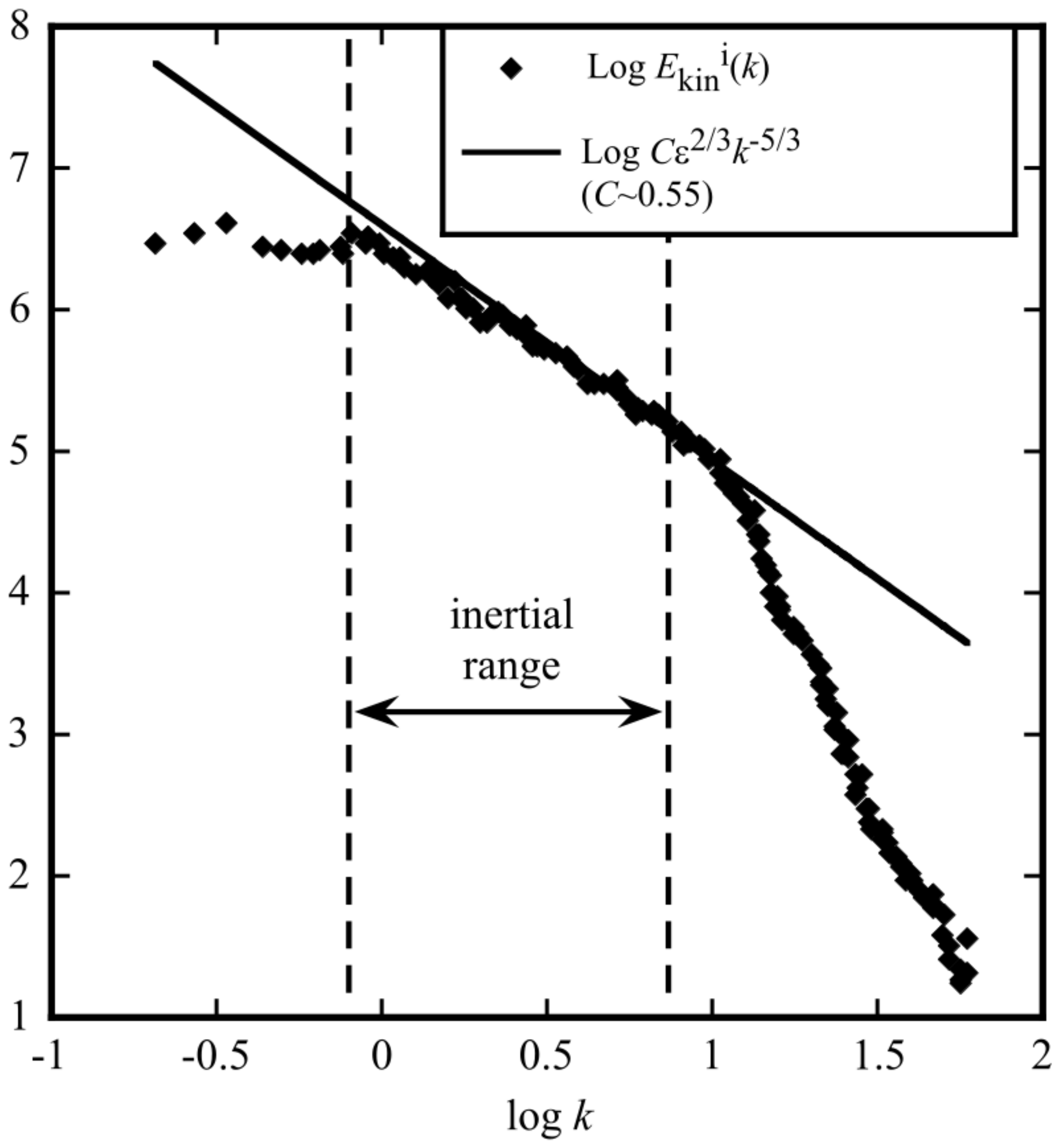} 
\caption{
Seminal work of Tsubota where the incompressible kinetic energy spectrum for a turbulent state is computed using the GPE.
The solid line corresponds to the Kolmogorov law.
Figure reprinted from M. Tsubota, “Quantum turbulence—from
superfluid helium to atomic Bose–Einstein condensates,” J. Phys.
Condens. Matter 21, 164207 (2009). © IOP Publishing. Reproduced
with permission. All rights reserved.
}
\label{fig:tsubota09}
\end{figure}

The number of quantized vortices in BEC systems is much smaller than the one observed in liquid helium experiments.
The typical number of atoms in BECs is close to a few hundred thousand, while liquid helium experiments are carried out with macroscopic samples.
Also, the ratio $L/\xi$ of the typical largest scale of the system $L$ to the healing length $\xi$ quantifies the range available for turbulence to take place. For superfluid He this value is $\log(L/\xi)\sim 4$, while for BECs $\log(L/\xi)\sim 1$ or 2. This means that the intrinsic spatial limitation in confined BECs prevents large-scale self-similar structures \cite{tsatos16}.
A natural question arises at this point: whether or not
Kolmogorov's scaling appears if the range of length scales available is so small.
Numerical simulations \cite{berloff02,kobayashi07,tsubota09,nowak11}
obtained a kinetic energy spectrum consistent with the $k^{-5/3}$ Kolmogorov scaling, see Fig.~\ref{fig:tsubota09}.

The attempts to model the transition to the turbulent state have been carried out in several works.
In Ref.~\cite{shiozaki11}, the authors characterized the transition using a critical number of vortices,
sample size, and
external energy input.
The perturbed system was classified by a phase diagram \cite{yukalov14,yukalov15}, and the granular phase was investigated \cite{yukalov14}.

Techniques to directly resolve the structure of individual vortices in experiments \cite{bretin03,fonda14,serafini15,serafini17}
open up the possibility of studying the
dynamics of a turbulent vortex tangle, which would be of paramount importance to the understanding of quantum turbulence.


Experimental investigations of turbulent BECs usually hold atoms
for tens of milliseconds in the trap, and then release the trap for imagining. A time of flight (TOF) absorption technique is used to obtain an image of the cloud.
The resulting image is a distorted two-dimensional projection of the spatial atomic distribution.
As a result, the spatial density has contributions from several wave numbers along the direction of the imaging light.
The turbulent regime is kinetically dominated, hence the interaction energy is assumed to be negligible.
The ballistically expanding atoms \cite{dalfovo99} allow for an experimental Fourier transform of the real space density distribution after a TOF to a momentum distribution.
After a time $t_{\rm TOF}$, the distance that an atom has traveled from the center of the trap is given by $r=\hbar t_{\rm TOF}k/m$.
The spatial distribution of atoms in free expansion can be used to map the momentum distribution, $n(r)\sim n(\hbar t_{\rm TOF}k/m)$, and the
momenta of the atoms are connected with the position in the expanded cloud. 
The radii of the expanded cloud are converted in momentum shells, and the number of atoms are counted in each shell to construct the momentum distribution of the sample.

In Ref.~\cite{thompson13}, the TOF technique was used to obtain the momentum distribution of a turbulent BEC cloud.
They found a power-law behavior for the momentum distribution
of $n(k)\propto k^{-2.9}$.
This scaling behavior in the momentum distribution, although probably affected by the presence of vortices, is consistent with the coexistence of yet another type of out-of-equilibrium dynamics led by nonlinear interaction of density waves (see Sec.~\ref{sec:wave}). Subsequent experiments have shown similar features in a box-trapped BEC \cite{navon16}. The momentum distribution analysis can be done in conjunction with particle and energy fluxes in order to characterize cascade processes \cite{navon19,orozco20,Marino2020}.

\subsection{2DQT}\label{sec:2DQT}

\begin{figure}
	\begin{center}
		\includegraphics[width=0.35\textwidth]{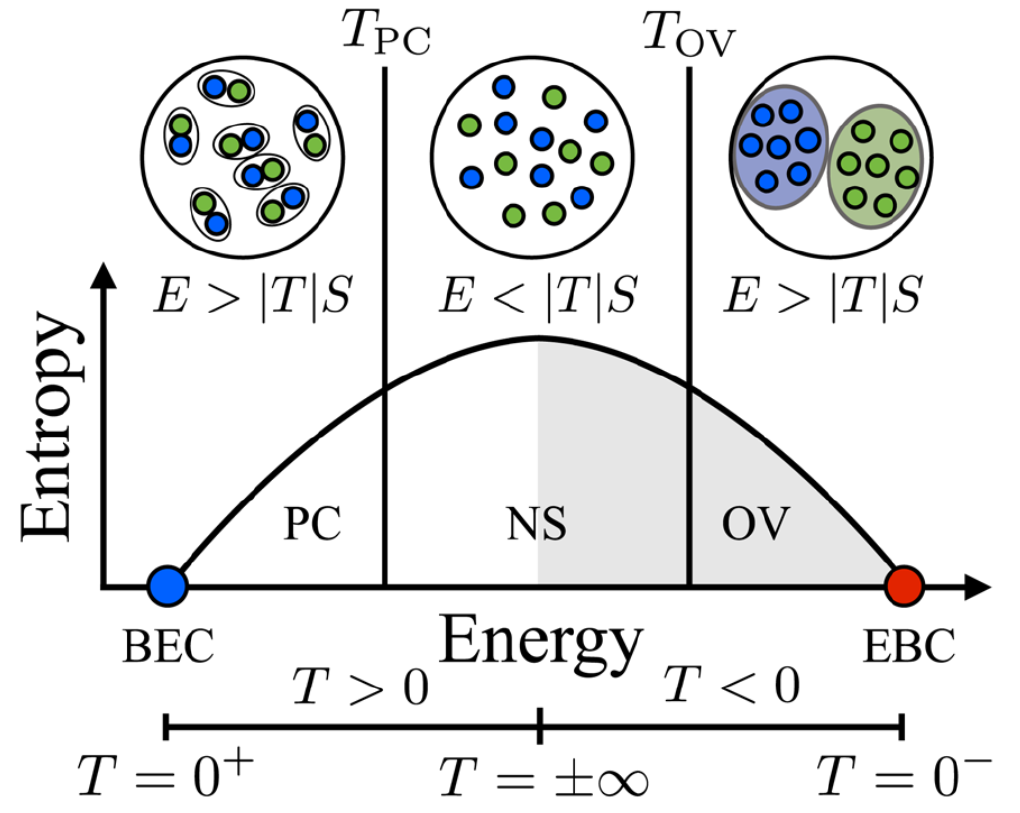} 
		\caption{Negative-temperature and Onsager condensation. The scheme shows the behavior of entropy for a point vortex model as a function of temperature. As temperature decreases and reaches negative values, the system passes through a phase-transition where vortex clustering, a coherent Onsager vortex (OV) state, becomes favorable against vortex dipoles and unbound vortices. At $T = \pm \infty$ entropy is a maximum and the vortex distribution is said to be in a entropy dominated normal state (NS). The vortex binding-unbinding phase transition separates this normal state from the pair collapse (PC) state at positive temperature.
		Reprinted figure with permission from T. Simula, M. J. Davis, and K.
Helmerson, “Emergence of order from turbulence in an isolated
planar superfluid,” Phys. Rev. Lett. 113, 165302 (2014). Copyright
2014 by the American Physical Society.}
		\label{Fig:negT}
	\end{center}
\end{figure}

Nowadays, experiments with BECs can achieve a true two-dimensional (2D) limit\cite{neely13,kwon14,seo17}, meaning that dynamics is completely frozen in one chosen direction. This implies that one-dimensional topological defects (vortex lines) become in practice points on a plane, with positive or negative circulation charges, a vortex or anti-vortex, respectively. Reconnection of vortex lines are replaced by vortex-antivortex annihilation processes, which release phonon-like excitations into the system \cite{neely13,kwon14,stagg16,cidrim16}.
It is also possible to confine the atoms in spherically symmetric potentials, known as bubble traps, which also produce 2D condensates, but the non-zero curvature introduces interesting aspects \cite{garraway16,padavic17,bereta19,tononi19,moller20}.
The impact of dimensionality in QT can also be investigated by studying a transition from 3D to 2D systems. For instance, a recent study showed a critical-like transition by numerically solving the GPE \cite{muller20}.

Kolmogorov and Vinen regimes of turbulence are also applicable for 2D. However, due to the reduced dimensionality, the introduction of large-scale flow by vortex bundling is replaced by the analogous 2D process of \textit{vortex clustering}. As in the classical turbulence case, such clustering happens as a result of forcing the system in small scales (as opposed to 3D, where injection of energy happens in large scales). Related physics was predicted by Onsager \cite{eyink06} in the context of an idealized 2D \textit{vortex gas} (modeled as points) in statistical equilibrium. Interestingly, Onsager's classical theory found in the 2D BEC a quantum testbed \cite{simula14,billam14,groszek16,valani18}. In Onsager's theory, the clustering process represents a phase transition to a certain vortex configuration associated with a negative-temperature state (see Fig.~\ref{Fig:negT}). Such effective temperature $T$ can be defined in terms of the vortex-gas configurational entropy $S\equiv E/T$ with (incompressible kinetic) energy $E$. Such clusters are long-lived structures \cite{gauthier19}, in contrast with vortex dipoles -- a pair composed of vortex and anti-vortex, with high annihilation probability (see schematic plot in Fig.~\ref{Fig:negT}). This is known as the Onsager vortex condensation, a key signature of an inverse energy cascade in 2D turbulent systems, transferring (incompressible) energy from small to larger scales.  

\begin{figure}
	\begin{center}
		\includegraphics[width=0.45\textwidth]{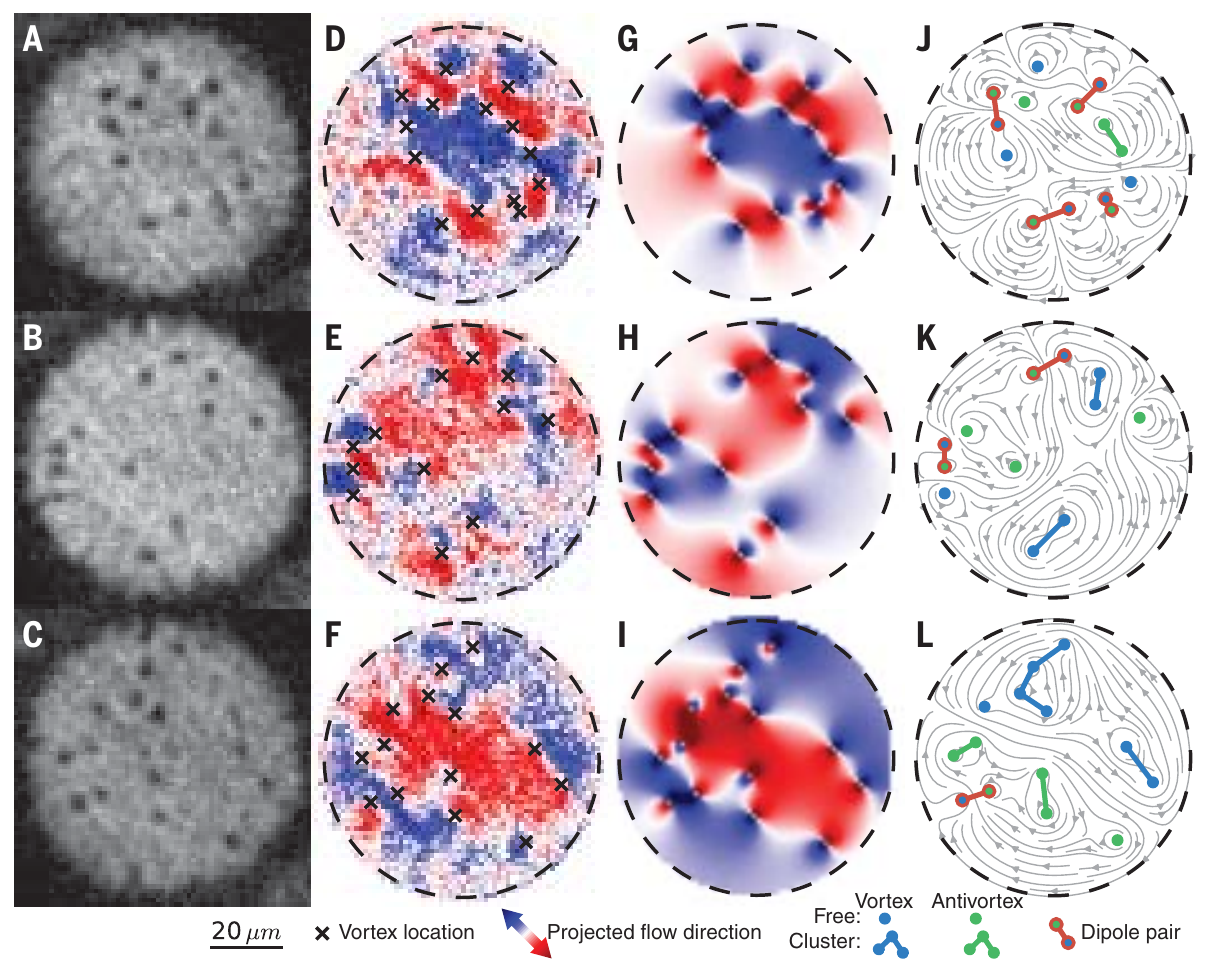} 
		\caption{ (A to C) Optical density images of experiment with perturbed BEC displaying different vortices (dark spots) configurations (dipole, random, and cluster
		respectively). (D to F) Corresponding Bragg spectroscopy signals, highlighting flow direction with different colors. (G to I) Associated numerically computed velocity field from point-vortex model. Colors in (D) to (I)
		indicate projections of the superfluid flow in the direction indicated by the arrow. (J to L) Formation of same-sign vortex clusters.
From S. P. Johnstone, A. J. Groszek, P. T. Starkey, C. J. Billington, T.
P. Simula, and K. Helmerson, “Evolution of large-scale flow from
turbulence in a two-dimensional superfluid,” Science 364,
1267–1271 (2019). Reprinted with permission from AAAS.
}  
	\label{fig:2DQT_johnstone}
\end{center}
\end{figure}

The ideas of Onsager have been successfully implemented in recent experiments \cite{gauthier19,johnstone19} (see Fig.\ref{fig:2DQT_johnstone}), where the emergence of negative-temperature vortex states indicate an inverse energy cascade.

\subsubsection{Kolmogorov turbulence in 2D}

\begin{figure}
	\begin{center}
		\includegraphics[width=0.35\textwidth]{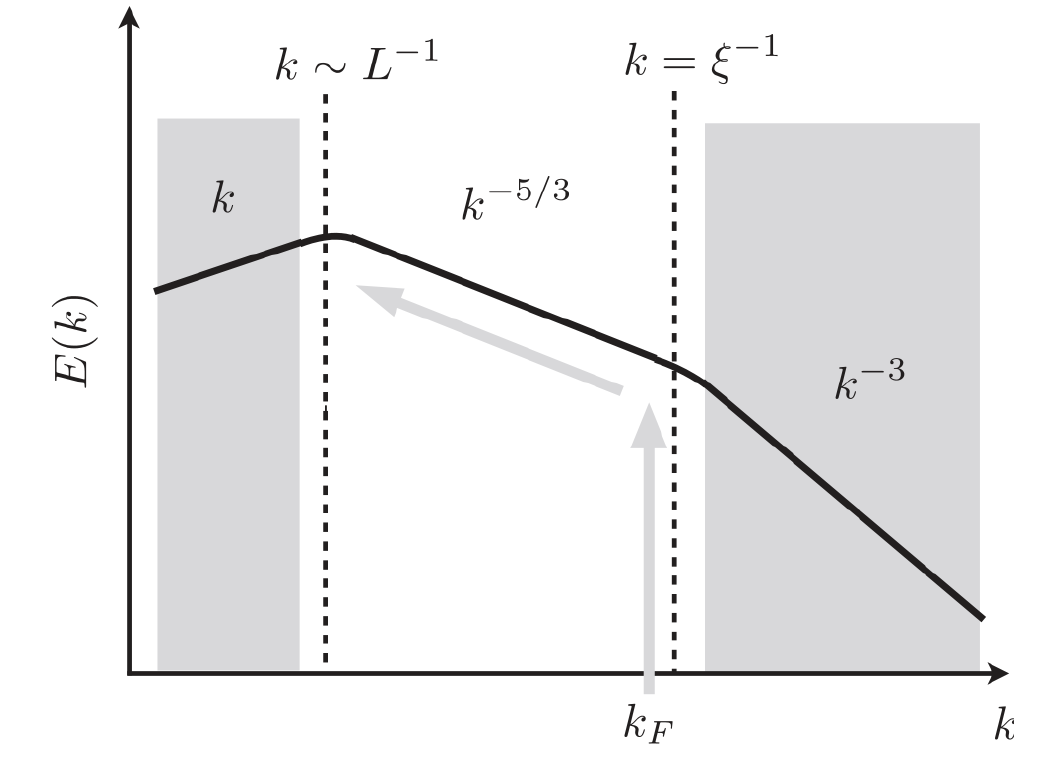} 
		\caption{Qualitative picture of the incompressible kinetic energy spectra for a 2D system. The $k^{-3}$ part of the spectrum appears due to the structure of the vortex core while the $k$ region pertains to distances larger than the largest intervortex distance $L$ and has no net vorticity. The nonshaded region is the inertial range where Kolmogorov scaling $k^{-5/3}$ manifests and $k_F$ is forcing scale, where energy is injected. 
Reprinted figure with permission from Ashton S. Bradley and Brian
P. Anderson, “Energy Spectra of Vortex Distributions in Two-
Dimensional Quantum Turbulence,” Phys. Rev. X 2, 041001 (2012).
Copyright 2012 by the American Physical Society.
}
		\label{2Dspectra}
	\end{center}
\end{figure}

Several numerical studies on 2D homogeneous superfluids~\cite{bradley12,reeves13,billam15} investigated the emergence of the quasiclassical regime, where the incompressible kinetic energy spectrum follows $k^{-5/3}$ (in 2D, actually known as Kraichnan-Kolmogorov spectrum \cite{kraichnan80}). Fig.~\ref{2Dspectra} shows a schematic version of the spectrum for a turbulent BEC forced in small scales $k_\mathrm{F}\sim \xi^{-1}$. The Kolmogorov scaling, appearing for large spatial scales, is associated with an inverse energy cascade through the vortex clustering, predicted by Onsager's vortex-gas theory. However, in 2D classical turbulence, the existence of a second inviscid quadratic invariant (besides the kinetic energy) -- the enstrophy, a measure of vorticity variance -- implies that a second, downscale cascade is present. This enstrophy transfer, accompanied by the upscale energy cascade, is expected to scale with $k^{-3}$. In superfluids, the quantized nature of circulation makes enstrophy proportional to the total number of vortices \cite{bradley12,white14} and hence the possibility of vortex-antivortex annihilations could typically force this quantity not to be an inviscid quantity in the quantum counterpart. For this reason, the conservation, and even the meaning, of enstrophy in quantum systems has been openly debated in the field \cite{numasato10, billam15}. Recently, however, the direct enstrophy cascade has been observed in simulations of a dissipative point-vortex model that described a two-dimensional, decaying turbulent quantum fluid~\cite{reeves17}. In this work, for a sufficiently large number of vortices ($\gtrsim 500$), the energy spectrum exhibited a clear enstrophy-related $k^{-3}$ scaling.  

\subsubsection{Vinen turbulence in 2D} 

Recent experimental and numerical efforts \cite{kwon14,kwon15,stagg15,cidrim16} have been focusing on understanding the decay of Vinen turbulence in 2D (and also aiming at an eventual extension of studying the decay in the Kolmogorov regime, still unexplored). Experiments with a sodium gas confined in a quasi-2D trap brought a BEC to a turbulent state by sweeping a repulsive laser beam of Gaussian shape through its center\cite{kwon14,seo17}. After producing $\sim 60$ vortices, the repulsive laser beam was turned off. The number of vortices $N$ as function of time was observed to decay non-exponentially. A phenomenological description (analogous to the 3D Vinen's equation \cite{vinen57}) was thus proposed by means of a rate equation given by
\begin{equation}
\frac{dN}{d t} = -\Gamma_1 N - \Gamma_2 N^{2},
\label{Eq:DecayKwon}
\end{equation}
with $\Gamma_1$ and $\Gamma_2$ real, positive parameters, attributed to one-body and two-body losses. This rate equation was in principle regarded to be universal. However, numerical simulations and phenomenological modeling have shown that vortex annihilation may involve leading four-body processes \cite{schole12,cidrim16,groszek16} in the zero-temperature case. Recent results on finite-temperature modeling suggest that different dissipative regimes induce other leading processes instead, such as three-body encounters \cite{groszek20}. 

\subsection{Wave turbulence}
\label{sec:wave}

Up to this point, we have mainly considered the Kolmogorov and Vinen turbulences, which are associated with chaotic configurations of vortex lines. Now we are going to consider the situations in which the chaotic behavior is associated with the condensate wave function itself. This is the so called \textit{wave turbulence} (WT). In general a system described by a complex wave function in Fourier representation $\psi(\mathbf{k})= A e^{i \theta}$ can be split into its real amplitude $A(\mathbf{k})$ and phase $\theta(\mathbf{k})$. The system will then be considered to be in a wave-turbulent regime if $\theta(\mathbf{k})$ for different values of $\mathbf{k}$ are statistically independent and uniformly distributed between $0$ and $2\pi$. In such conditions, it turns out that the nonlinear interaction between the different Fourier modes of the wave function can also give rise to self-similar cascades which can be observed through power laws in the energy spectrum \cite{fujimoto15}. In the case where the kinetic component of the energy dominates over the interaction component, an elegant analytical treatment is possible \cite{nazarenko11book, zakharov92book} in which is know as \textit{weak wave turbulence}. 

One remarkable feature in the theory is the fact that  properties associated with stationary turbulent states can be directly obtained from the wave dispersion relation and the lowest-order nonlinear term. In BECs there are two main types of wave turbulence: four-wave and three wave turbulences.
For the sake of simplicity, we consider here an infinite system with no external trapping potential. 

\subsubsection{Four-Wave turbulence}
As a first example of WT let us consider a very dilute BEC at $T=0$ which can be described by the dimensionless version of GPE in Fourier representation 

\begin{align}
i\frac{\partial \psi}{\partial t} = \omega(k) \psi + g^{\prime} & \int d\mathbf{k}_1 \int d\mathbf{k}_2 \int d\mathbf{k}_3  \delta(\mathbf{k}-\mathbf{k}_1-\mathbf{k}_2+\mathbf{k}_3) \nonumber \\ &\times V(\mathbf{k}, \mathbf{k}_1,\mathbf{k}_2,\mathbf{k}_3) \psi(\mathbf{k}_1)\psi(\mathbf{k}_2)\psi^{\ast}(\mathbf{k}_3) ,
\label{Eq:GPE-ad}
\end{align}	
where $\omega(k)=k^2 / 2$ and $V(\mathbf{k}, \mathbf{k}_1,\mathbf{k}_2,\mathbf{k}_3)=1$.  Such a four-wave weakly interacting system in the WT regime must be described by following equation for the wave action \cite{nazarenko11book} $n(\mathbf{k})= \lvert \psi(\mathbf{k})\rvert^2$:

\begin{align}
	\frac{\partial n (\mathbf{k})}{\partial t}=& 4\pi g^{\prime 2} \int d\mathbf{k}_1 \int d\mathbf{k}_2 \int d\mathbf{k}_3  \delta(\mathbf{k}-\mathbf{k}_1-\mathbf{k}_2+\mathbf{k}_3) \nonumber \\ & \times \delta(\omega(\mathbf{k})-\omega(\mathbf{k}_1)-\omega(\mathbf{k}_2)+\omega(\mathbf{k}_3)) \nonumber \\ & \!\!\!\!\!\!\!\!\!\!\!\!\!\!\!\!\!\!\!\!\times n(\mathbf{k})n(\mathbf{k}_1)n(\mathbf{k}_2)n(\mathbf{k}_3)\left[ \frac{1}{n(\mathbf{k})} + \frac{1}{n(\mathbf{k}_3)}-\frac{1}{n(\mathbf{k}_1)}-\frac{1}{n(\mathbf{k}_2)}\right].
	\label{Eq:four-wave}
\end{align}	

Using Zakharov transformation on the $k$ variables \cite{zakharov92book} two power law solutions of the type $n \sim k^\nu$ are possible in this system with
\begin{align}
	\nu_E &= -2\beta/3 -d, \\
	\nu_N &= -2\beta/3 -d +\alpha/3,
	\label{Eq:four-wave-sol}
\end{align}	
which correspond to energy ($\nu_E$) and particle ($\nu_N$) cascades. Here $d$ is the spatial dimension of the system while $\alpha$ and $\beta$ are the degree of homogeneity of $\omega$ and $V$, i.e., $\omega(\lambda k) = \lambda^\alpha \omega(k)$ and $V(\lambda\mathbf{k}_1,\lambda\mathbf{k}_2,\lambda\mathbf{k}_3,\lambda\mathbf{k}_4) = \lambda^\beta V(\mathbf{k}_1,\mathbf{k}_2,\mathbf{k}_3,\mathbf{k}_4)$. In the case of 3D GPE, we have $\alpha = 2$, $\beta=0$ and $d=3$. These parameters lead to the energy and wave action cascades power law spectrum which are $\nu_E = -3$ and $\nu_N = -7/3$, respectively.

\subsubsection{Three-Wave turbulence}

In BECs, this is the turbulence associated to the small amplitude excitations $\delta\psi$ over a macroscopic background \cite{kevrekidis08book} $\psi_0$. The equation of motion for such excitations can be obtained by substituting the Ansatz $\psi=\left[\sqrt{\rho_0} + \delta\psi \right] e^{-ig^{\prime}\rho_0 t}$ and keeping only the smallest-order nonlinearities 

\begin{equation}
	i \frac{\partial\delta\psi}{\partial t}=-\frac{1}{2}\nabla^2 \delta\psi + g^\prime \rho_0 (\delta\psi +\delta\psi^\ast)+ g^\prime \sqrt{\rho_0}(2 \delta\psi^\ast \delta\psi + \delta \psi^2),
	\label{Eq:phonons}
\end{equation}	
which has the Fourier representation

\begin{align}
	i \frac{\partial \delta\psi (\mathbf{k})}{\partial t}&= \frac{k^2}{2}\delta\psi(\mathbf{k}) + g^\prime \rho_0\left[ \delta\psi(\mathbf{k}) +\delta\psi^\ast (\mathbf{k})\right] \nonumber \\ & +\frac{g^\prime \rho_0}{(2\pi)^{3/2}} \int d\mathbf{k}_1\int d\mathbf{k}_2 \left[ \delta(\mathbf{k} - \mathbf{k}_1 -\mathbf{k}_2)\delta\psi(\mathbf{k}_1)\delta\psi(\mathbf{k}_2) \right. \nonumber \\ & \qquad\qquad \left. + 2  \delta(\mathbf{k} + \mathbf{k}_1 -\mathbf{k}_2)\delta\psi^\ast (\mathbf{k}_1)\delta\psi(\mathbf{k}_2)  \right]
	\label{Eq:phonons-k}
\end{align}

The Bogoliubov transformation \cite{lvov03,dyachenko92} at this point is necessary in order to make explicit the dynamics of plane waves

\begin{align}
	a(\mathbf{k}) =& \frac{1}{2} \left( \sqrt{\frac{\omega(k)}{\beta(k)}}+\sqrt{\frac{\beta(k)}{\omega(k)}}\right)\delta\psi(\mathbf{k}) \nonumber \\ &+\frac{1}{2} \left( \sqrt{\frac{\omega(k)}{\beta(k)}}-\sqrt{\frac{\beta(k)}{\omega(k)}}\right)\delta\psi^\ast (-\mathbf{k}),
\end{align}
with its inverse transformation
\begin{align}
	\delta\psi(\mathbf{k}) =& \frac{1}{2} \left( \sqrt{\frac{\omega(k)}{\beta(k)}}+\sqrt{\frac{\beta(k)}{\omega(k)}}\right)a(\mathbf{k}) \nonumber \\ &-\frac{1}{2} \left( \sqrt{\frac{\omega(k)}{\beta(k)}}-\sqrt{\frac{\beta(k)}{\omega(k)}}\right) a^\ast (-\mathbf{k}),
\end{align}
where $\beta(k)=k^2 /2$ and $\omega(k)=\sqrt{\beta(k)^2 + 2g^\prime \rho_0 \beta(k)}$. Thus, substituting into (\ref{Eq:phonons-k}) we get the equation for weakly interacting sound waves
\begin{align}
	i \frac{\partial a (\mathbf{k})}{\partial t}&= \omega(k) \delta\psi(\mathbf{k}) \nonumber \\ & + \int d\mathbf{k}_1\int d\mathbf{k}_2 V(\mathbf{k};\mathbf{k}_1,\mathbf{k}_2 ) \left[ \delta(\mathbf{k} - \mathbf{k}_1 -\mathbf{k}_2)a(\mathbf{k}_1)a(\mathbf{k}_2) \right. \nonumber \\ & \qquad\qquad \left. + 2  \delta(\mathbf{k} + \mathbf{k}_1 -\mathbf{k}_2 )a^\ast (\mathbf{k}_1)a(\mathbf{k}_2)  \right],
\end{align}
where
\begin{align}
V(\mathbf{k};\mathbf{k}_1,\mathbf{k}_2 ) =& \frac{\sqrt{g^{\prime}\rho_0\omega(k_1)\omega(k_2)\omega(k_3)}}{(2\pi)^{d/2}}\left[ \frac{6}{\alpha(k_1)\alpha(k_2)\alpha(k_3)} \right. \nonumber \\	& \left. + \frac{1}{2}\left( \frac{\mathbf{k}_1\cdot\mathbf{k}_2}{k_1k_2\alpha(k_3)}
+\frac{\mathbf{k}_2\cdot\mathbf{k}_3}{k_2k_3\alpha(k_1)}
+\frac{\mathbf{k}_3\cdot\mathbf{k}_1}{k_3k_1\alpha(k_2)}
\right) \right],
\end{align}
and $\alpha(k)=2 g^{\prime}\rho_0+ k^2$. 

If $\delta\psi$ follows the WT conditions then the equation for the wave action $n(\mathbf{k})= \lvert \psi(\mathbf{k})\rvert^2$ is \cite{nazarenko11book}

\begin{equation}
	\frac{\partial n (\mathbf{k})}{\partial t}= \int d\mathbf{k}_1 \int d\mathbf{k}_2 \left(
	R(\mathbf{k}_1,\mathbf{k}_2,\mathbf{k})-	R(\mathbf{k},\mathbf{k}_1,\mathbf{k}_2)-
	R(\mathbf{k}_2,\mathbf{k},\mathbf{k}_1)
	\right)
	\label{Eq:three-wave}
\end{equation}	
where 
\begin{align}
R(\mathbf{k}_1,\mathbf{k}_2;\mathbf{k}_3)=&2\pi \lvert V(\mathbf{k}_3;\mathbf{k}_1,\mathbf{k}_2 ) \rvert^2 \delta(\mathbf{k}_3-\mathbf{k}_1-\mathbf{k}_2) \nonumber \\ & \times \delta(\omega(k_3)-\omega(k_1)-\omega(k_2)) \nonumber \\  & \times \left[ n(\mathbf{k}_1)n(\mathbf{k}_2)-n(\mathbf{k}_2)n(\mathbf{k}_3) -n(\mathbf{k}_3)n(\mathbf{k}_1) \right].
\end{align}
In the strong condensate limit $g^{\prime}\rho_0 \gg k^2$, the system simplifies so that
\begin{align}
\omega(k) & \sim k , \\
V(\mathbf{k};\mathbf{k}_1,\mathbf{k}_2 ) 
& \sim \sqrt{k_1 k_2 k_3},
\end{align}
therefore in such a limit, the degree of homogeneity of $\omega$ and $V$ are $\alpha =1$ and $\beta = 3/2$,  respectively. This leads \cite{nazarenko11book} to a cascade solution $n\sim k^\nu$ with $\nu = -\beta -d$ which in the 3D case gives $\nu = -9/2$.

\subsection{QT from the out-of-equilibrium perspective}

Many recent works have looked at the problem of quantum turbulence from a more general standpoint. A set of out-of-equilibrium phenomena can be categorized into classes of systems that present universal dynamical behavior. This is done in analogy to scaling theories for systems in equilibrium, but under the light of a renormalization-group theory that treats time as a scaling parameter when the system is far from equilibrium \cite{schmidt12, nowak14, orioli15, schmied19, chantesana19}. The universal dynamical behavior can be related to the presence of a so-called non-thermal fixed point, a metastable state of the perturbed quantum many-body system that can be universally characterized. It is quite possible that understanding the dynamics of isolated quantum systems away from steady state, as well as their quest for equilibrium, can elucidate many aspects of QT. It is believed that a large class of quantum systems outside equilibrium, including QT, has universal behavior in their temporal and spatial evolution. Experiments in this direction begin to generate results \cite{prufer18,erne18,Glidden2020}. In these cases, independently of the initial conditions, the system has dynamical evolution characterized by only a few parameters.

\section{Quantum turbulence in ``exotic'' systems}
\label{sec:exotic}

So far, we limited ourselves mainly to the discussion of single-component BECs (we should note that even in this case there are two elements, the thermal cloud and the superfluid \cite{tavares13}).
However, QT can also be studied in a plethora of systems.
We should stress that we present brief introductions to each of the topics listed in this section,
since there is enough material in the literature to write a complete review
about each of them.
We focus on the main aspects of each topic while presenting the features
that can be readily connected to the majority of studies of QT in
trapped single-component BECs.

\subsection{Bosonic mixtures}
\label{sec:mixture}

Quantum turbulence can be studied in a mixture of two, or possibly more,
bosonic species.
A combination widely employed is a Na-K mixture, due to the
relatively simple experimental procedure involved and considerable flexibility of the system \cite{castilho19}.

Studies began with a single vortex in multicomponent BECs \cite{kasamatsu05}.
The development of quantum turbulence from two counter-propagating superfluids of miscible Bose-Einstein condensates has been investigated
numerically by solving the coupled Gross–Pitaevskii equations \cite{takeuchi10,ishino11}.
This can be seen as the analog of quantum turbulence in $^4$He, in a regime where the normal and superfluid
components are turbulent at the same time.
It was found that
when the relative velocity exceeds a critical value, the counterflow becomes unstable, and quantized vortices and vortex rings are nucleated, which leads to isotropic quantum turbulence consisting of two superflows.

Another theoretical study \cite{kobyakov14}, also using numerical solutions
of the Gross-Pitaevskii equations, observed the Kolmogorov scaling law for the incompressible kinetic energy in a binary immiscible mixture of
$^{87}$Rb atoms.

A fascinating situation would be to employ one of the components
of the mixture as a probe for the other, using either
an impurity\cite{astrakharchik04} or a comparable number of atoms
of the second component.
One of the goals would be the
visualization of the vortex line tangle that constitutes
turbulence, which is very difficult in a trapped BEC, but it
is well-developed
in liquid He \cite{bewley06}.

\subsection{Spin turbulence}
\label{sec:spin}

Condensates with particles possessing a spin degree of freedom
have been produced \cite{kurn13}.
The study of turbulence in these systems, mainly spin-1 spinor
BECs, is called spin turbulence (ST).
The turbulent state is characterized by the spin density vectors having various disordered directions.
We should stress that, in the case of ST, turbulence is not referring to the
state of the mass density, but rather the spin density. A theoretical investigation or experiment probing both types of turbulence at the same time would be very challenging.

One of the main differences between ST and a regular mixture of
two bosonic components are that spin-exchange collisions make the
number of atoms of each component fluctuates in spin systems, whereas the populations
are constant in spinless BECs.
A hydrodynamical description of ST is available \cite{fujimoto12,fujimoto12_2}.

The properties of ST depend on whether the spin-dependent interactions are ferromagnetic or antiferromagnetic.
In theoretical studies \cite{fujimoto12,fujimoto12_2,tsubota13}, the authors
observed a spectrum of the spin-dependent ferromagnetic interaction energy displaying a -7/3 power-law, which is different from the -5/3 Kolmogorov scaling.
Also, ST can behave as a spin-glass, corresponding to random spin density vectors, but frozen in time \cite{tsubota13}.

Besides spin-1 BECs,
numerical calculations of a spin-2 BEC suggest that QT is possible in these systems \cite{mawson15}.
An exciting experiment exploring 
the interaction and dynamics of half-quantum vortices in an antiferromagnetic spinor Bose-Einstein condensate showed that turbulent condensates emerge from many half-quantum vortices collisions \cite{seo16}.
For a review of the theoretical and numerical works on ST, the reader is referred to Ref.~\cite{tsubota14}.

\subsection{Fermionic gases}
\label{sec:fermionic}

So far, we have only discussed superfluids containing bosonic constituents.
However,
superfluidity is also possible in fermionic systems.
The Bardeen-Cooper-Schrieffer (BCS) theory of condensation of Cooper pairs
into bosonic-like particles \cite{bardeen57} explains the mechanism behind
fermionic superfluids.

Interatomic interactions can be tuned, both in bosonic and fermionic
dilute gases, using Feshbach resonances. Bosonic systems with attractive pair-wise interactions will eventually
collapse \cite{sackett99}. However, that is not the case with fermions
due to the Pauli exclusion principle.
The interparticle interactions
can be tuned so that the fermion pairs change their size from
tightly bound dimers (BEC limit) to many times the interparticle distance
at the BCS side, spanning the so-called BEC-BCS crossover \cite{randeria14,strinati18}. At the center of the crossover, there is
the strongly-interacting unitary regime, with remarkable properties.
A milestone for the study of trapped fermionic superfluids was the observation of vortex lattices throughout the BEC-BCS
crossover in a $^6$Li gas, which demonstrates superfluidity
\cite{zwierlein05}.

Quantum turbulence is possible in fermionic gases, at least
in the unitary regime \cite{wlazlowski15,wlazlowski15_2}.
Naturally, many fundamental questions arise.
In which portions of the BEC-BCS crossover it is
possible to observe QT?
What is the impact of the quantum statistics - bosons versus fermions -
in turbulence?
Although the answers seem far away,
the microscopic structure of vortices in cold atomic fermionic gases has been studied throughout the BEC-BCS
crossover and in the unitary Fermi gas \cite{mad16,mad17,madeira19}, and
the time-dependent superfluid local density approximation
is rather reliable to obtain static and dynamic phenomena \cite{bulgac16}.

\subsection{Astrophysics}
\label{sec:astrophysics}

\subsubsection{Neutron stars}

Quantum turbulence may provide answers to a mystery in nuclear astrophysics:
the pulsar glitches. These are sudden increases in the spinning
of neutron stars, while they continue to lose angular momentum.
Since neutrons are spin-1/2 particles, QT of fermionic
gases, see Sec.~\ref{sec:fermionic}, is of interest.
One possible explanation is that the
outer core of a neutron star is in a turbulent state and that
the Reynolds number could account for the glitches
\cite{peralta06,link13}.

As is the case with turbulence in trapped BECs,
this problem would benefit from a better understanding of microscopic processes, such as vortex reconnections.
However,
even the study of a single straight vortex line in neutron matter
is an
active topic of research \cite{Elgaroy2001,yu03,madeira19}.

One of the main challenges in this problem is the range of available length scales. Although this is true for all problems related to turbulence, this feature is extreme in the case of neutrons stars.
The vortex cores are in the femtometer scale, while neutron stars are in
the kilometer scale.
Some
progress has been made
toward developing a mean-field framework for this situation \cite{khomenko18}.

\subsubsection{Turbulence in magnetohydrodynamics}

Magnetohydrodynamics (MHD) is the study of magnetic properties of electrically conducting fluids; such as the case of plasmas.
Turbulence in plasmas seems to be far away from the turbulent
regime of trapped BECs.
However, they both share common aspects.
For example, there is wave turbulence \cite{cho03} in MHD,
magnetic reconnections \cite{lazarian99} are present,
and ongoing debates about the power-law exponent of the energy spectrum
\cite{perez12}.
Turbulence in MHD and QT in trapped BECs also present
an anisotropic aspect, and understanding some concepts of turbulence in MHD may prove useful to improve the description
of turbulence in BECs.

Anisotropy arises in BECs due to the trap, which is evident in the
commonly employed cigar-shaped clouds.
In order to understand the origin of the anisotropy in MHD, let us
start with a static homogeneous plasma \cite{schekochihin19}.
We can think of it as describing a local portion of a much larger system.
Even if the equilibrium quantities such as density and pressure are
large-scale, the only large-scale feature that does not vanish at
small scales is the magnetic field, hence defining a preferential direction.
This is what makes MHD turbulence different from rotating or stratified turbulence, which always reverts to the universal Kolmogorov distribution at small enough length-scales \cite{nazarenko11}.

Kraichnan \cite{kraichnan65} used this irreducibility of the magnetic field
to derive an energy spectrum. The background uniform magnetic field
$\textbf{B}_0$ can be converted into velocity units, the so-called Alv\'en
speed, $v_A=B_0/\sqrt{4\pi\rho_0}$, $\rho_0$ being the mass density of the medium.
Then, by dimensional analysis, the energy spectrum in the inertial range must be
\begin{equation}
E(k) \propto (\epsilon v_A)^{1/2}k^{-3/2},
\end{equation}
which is known as the Iroshnikov–Kraichnan spectrum \cite{iroshnikov63}.
Kraichnan's interpretation of the spectrum was wrong, because
of the Kolmogorov assumption of restoration
of symmetry at small enough length scales, which leads to only one $k$
in the dimensional analysis.
In fact, there are two relevant wave numbers, $k_{\parallel}$
and $k_{\perp}$, representing the turbulent fields along and across $\textbf{B}_0$, respectively.

In a strong magnetic field, perturbations with $k_{\perp} \gg k_{\parallel}$
should happen more often than isotropic ones, because the
magnetic field lines are hard to bend.
This intuitive argument turns out to be right for the anisotropy of MHD turbulence at all length scales \cite{montgomery81}.
Goldreich and Sridhar put forward arguments and assumptions \cite{goldreich95,goldreich97}
to derive an energy spectrum of the form
\begin{equation}
E(k_\perp)\propto \varepsilon^{2/3} k_\perp^{-5/3},
\end{equation}
which is the anisotropic version of the Kolmogorov
scaling.

This very brief chronological exposition of events in the field of turbulence in MHD shows the importance of taking into account the anisotropic aspect of the system.
Perhaps, analogous directions could be pursued in QT in trapped BECs to
improve our description of these systems.

\section{Challenges ahead}
\label{sec:challenges}

The production of a turbulent regime in trapped condensates is a very recent topic of investigation.  Despite turbulence itself being a centenary theme, and turbulence in quantum fluids already being observed a few decades ago, our mathematical capacity as well as our models applicable to turbulence are still few. Most of the time, studies are trying to compare classical and quantum turbulence. Surprises are, however, very much expected in quantum fluids. There are, therefore, great challenges that we must face in this field.

\textbf{Defining the undefined} - one of the challenges for quantum turbulence is its definition. While for classical turbulence there are more consolidated criteria and interpretations for the definition of turbulence, in the quantum world, this has not yet reached maturity.
Today, most definitions of quantum turbulence are broad and undefined.
An analogy with the classical fluids is sought to reach some criteria that allow establishing whether a fluid is experiencing the turbulence or not. However, this does not seem to be the most appropriate way to proceed. Not everything in the quantum world has a counterpart to be compared to the classic one. In the classical world, it is conventional to classify turbulence in terms of the Reynolds number, focusing on viscosity. Increased viscosity makes the fluid less accessible to turbulence. There is no equivalent in the quantum world.
While we deal with a wide range of scales of motion in the classical world, we also have to deal with a wide range of scales capable of holding energy in the quantum world. This parallel, should allow an equivalent of the Reynolds number, with quantum characteristics. But this does not exist yet.
Although we have fundamental differences between classical and quantum fluids, experiments and theory have shown many similarities between classical and quantum turbulence. Several experiments on liquid helium show striking similarities with the observed behavior in classic fluids. However, for Bose-Einstein condensates confined in a trap, we focus on similarities, but in practice, great difficulties in making analogies with the classical world have been faced.
Perhaps the way to consider the case of trapped quantum fluids, is to consider them as a special class of quantum fluids as well as their turbulence. 
As it is typical in the quantum systems, it would be interesting to have a series of equations that governed average values of quantum quantities like momentum and energy. It could be possible to define turbulence criteria based on such large averages in instead of particular values. This would be a statistical model that would allow coming up with definitions for the turbulent regime. This, however, seems unrealistic to many. The possibility of describing quantum turbulence as a state of the physical system could be more realistic. The continuous search for quantum-classical analogies is a way of seeking to observe a more simplified quantum world and export concepts to the confused turbulent classical world. This can be an interesting way, but not necessarily valid.

\textbf{The issue of isotropy in condensates and turbulence} - While for simplicity, many models in turbulence are devised considering isotropic systems, where it is evenly distributed in all directions, this is in fact not a necessity and there are no restrictions to consider a different scenario. In condensates, the density is not constant and in most situations not spherically symmetric, depending on the type of confining potential. Moreover, the potential is almost never isotropic. In fact, in many experiments the system even has a large difference in values in the confinement of orthogonal directions. Allied to such anisotropic confinement is the finite character of the system, not allowing to create an unrestricted number of excitations, having limits, which can be different, in each of the directions. This makes the system intrinsically anisotropic and can transmit such properties to the turbulent cloud. The result of this situation is that one can have anisotropic characteristics, either in stationary or transient regime. These facts need to be explored, since without an isotropic system, the transfer and thermalizing  of excitations between directions can be a new and extremely relevant physics for the investigation of turbulence in  condensates.

\textbf{Turbulence formation} - One of the important points in the formation of turbulence involving Bose-Einstein condensates is the introduction of excitations (vortices or other density disturbances), which later evolve to the turbulent state. This process starts with a condensate in equilibrium, and then taking it out of equilibrium. The process is opposite to that elaborated by Kibble-Zurek \cite{lamporesi13}. This allows us to imagine that the transformation of a condensate in equilibrium to a turbulent one follows an inverse route to that of the Kibble-Zurek mechanism \cite{yukalov15_2}. The inverse Kibble-Zurek process can be a universal way to start with a trapped condensate and lead it to turbulence, with an evolution sequence between these extremes. Understanding this situation is certainly one of the challenges of the field.

\textbf{The out of equilibrium aspects} -Turbulence is one of the most intriguing phenomena of systems taken out of the equilibrium.
While we use concepts typical of classical fluids to investigate quantum turbulence,
looking for the decay aspects of an out of equilibrium quantum system is unique,
and may be the most correct way to treat such systems. With that in mind, we can investigate the universal critical behavior of the system, as a characteristic of a phase transition out of equilibrium. There are several ways to investigate the evolution of turbulence following those lines. One of the most interesting ways, being the so-called transition to absorptive states. Usually, it is a state where the system enters and can no longer escape from  it. Aspects of universality for systems out of equilibrium can be seen as the system gets in a non-thermal state that fixes itself around certain conditions (non-thermal fixed points).
In these systems, a universal dynamics can occur during the temporal evolution of the system, as recently demonstrated in quenched Bose gases \cite{erne18,Glidden2020}. In these cases, variables such as the momentum distribution can evolve over time, generating a set of exponents that allow a re-scaling within certain universal functions. Addressing quantum turbulence, from a perspective of universal dynamic in far from equilibrium system, can lead to a deeper and less empirical understanding of the turbulent regime.

\begin{acknowledgments}
This work was supported by
the S\~ao Paulo Research Foundation (FAPESP)
under the grants 2013/07276-1, 2014/50857-8, and 2018/09191-7, and by the
National Council for Scientific and Technological Development (CNPq)
under the grant 465360/2014-9. FEAS thanks CNPq for support through Bolsa de
produtividade em Pesquisa Grant No.
305586/2017-3.
\end{acknowledgments}

\section*{Data Availability Statement}

Data sharing is not applicable to this article as no new data were created or analyzed in this study.

\bibliography{aipsamp}

\end{document}